\documentclass[12pt]{iopart}
\usepackage{iopams}  
\begin{document}

\title [Optimal control theory for quantum-classical systems] {Optimal
  control theory for quantum-classical systems: Ehrenfest Molecular
  Dynamics based on time-dependent density-functional theory}

\author{A. Castro}
\address{
ARAID Foundation and Institute for Biocomputation and Physics of Complex Systems,
University of Zaragoza, Mariano Esquillor s/n, 50018 Zaragoza (Spain)
}
\ead{acastro@bifi.es}

\author{E. K. U. Gross}
\address{Max-Planck Institut f{\"{u}}r Mikrostrukturphysik, Weinberg 2,
D-06120 Halle, Germany.}

\begin{abstract}
  We derive the fundamental equations of an optimal control theory for
  systems containing both quantum electrons and classical ions. The
  system is modeled with Ehrenfest dynamics, a non-adiabatic variant
  of molecular dynamics. The general formulation, that needs the fully
  correlated many-electron wave function, can be simplified by making
  use of time-dependent density-functional theory. In this case, the
  optimal control equations require some modifications that we will
  provide.  The abstract general formulation is complemented with the
  simple example of the H$_2^+$ molecule in the presence of a laser
  field.
\end{abstract}

\submitto{\JPA}

\section{Introduction}

Molecular dynamics (MD)~\cite{Allen1989,Rapaport2004} is the field of
physical modelling dedicated to atomistic simulations of condensed
matter systems. Due to the impossibility of a full quantum treatment
for all particles, the nuclei are considered to be classical, whereas
the electrons must retain their quantum nature.  This classical
description of the nuclear system is of course an approximation, and
in many circumstances it is necessary to employ nuclear wave packets
to study the dynamics of molecules and other condensed matter
systems. The term ``molecular dynamics'', however, is traditionally
reserved for the models in which the nuclei are fully
classical. Therefore, the problem addressed by MD is that of the
evolution of a mixed system composed of a classical and a quantum
subsystem.

The level of theory used to describe the electrons differs in the various MD
schemes, ranging from detailed first principles
approaches~\cite{Marx2009}, to the so-called ``classical'' MD force
fields~\cite{Gavezzotti2007,Leach2001,Mackerell2004}, in which the electronic
subsystem is in fact integrated out, and the remaining ions interact with
classical forces that have been carefully parametrized over the years to
somehow contain the lost electronic influence.  In this latter case, the only
equations to follow are Newton's laws for the nuclei, and, at least
operationally, there is no longer a mixed system -- though, originally,
the full system is mixed.

Another broad division in MD can be established between adiabatic and
non-adiabatic models. In the former, the electronic system is considered to
be, at all times, at the ground state (or, perhaps, at some fixed excited
state) corresponding to the instataneous nuclear configuration. In
non-adiabatic MD, transitions between electronic states are allowed. These are
the cases that, strictly speaking, necessitate a genuine mixed
quantum-classical approach. Not surprisingly, the problems that require a
non-adiabatic model are computationally the most challenging, since they
require an ab-initio electronic structure model.

Yet another classification of MD studies can be established with respect to
the properties of the system that one wishes to study via the
simulations. Originally, the objects of investigation were assumed to be the
equilibrium properties in the canonical ensemble of macroscopic systems.
However, the out-of-equilibrium dynamics of molecules in the presence of high
intensity fields has become of enormous interest in the last decades due to
the experimental advances in that area. When laser pulses are to be
considered, one needs a first-principles non-adiabatic model, especially if
the pulses have high intensities and the frequencies are visible or higher,
i.e. of the order of typical electronic excitations.

In this work, we are interested in non-equilibrium laser-matter interaction
experiments, that require a non-adiabatic first principles model.  Ehrenfest
dynamics is one of such models. It can be derived by taking two consecutive
approximations: first, an electronic-nuclear separation of the full quantum
wave function leads to the usually called ``time-dependent self-consistent
field'' model~\cite{Gerber1982}; then, the short wave asymptotics of Wentzel,
Kramers and Brillouin (WKB)~\cite{Wentzel1926,Kramers1926,Bri1louin1926} is
used to take the classical limit for the nuclear degrees of freedom. A
discussion on its justification and validity can be found in
Ref.~\cite{Bornemann1996}. The term ``Ehrenfest dynamics'' is not of universal
use -- for example, in this aforementioned work of Bornemann \emph{et
  al.}~\cite{Bornemann1996}, the scheme is simply called ``mixed
quantum-classical dynamics''. The use of Ehrenfest's name is due to the fact
that the classical equation of motion for the nuclei can be obtained as an
application of his famous theorem~\cite{Ehrenfest1927}.

Since, for practical implementations, the many-electron wave function cannot
be directly handled, it is necessary to model the quantum dynamics with some
electronic structure theory, such as for example time-dependent
density-functional theory (TDDFT)~\cite{Runge1984,tddft}. Ehrenfest MD based
on TDDFT was first attempted by Theilhaber~\cite{Theilhaber1992} for
(external) field-free problems, and afterwards its utility to laser-matter
irradiation has been repeatedly demonstrated -- see,
e.g. Refs.~\cite{Saalmann1996,Calvayrac1998a,Saalmann1998,Kunert2001,Castro2004a,Kunert2005,Castro2012}.
Note, however, that TDDFT is, in practice, only an approximate theory (as some
of its ingredients -- most notably the ``exchange and correlation functional''
-- are unknown), and the range of applicability of the state-of-the-art TDDFT
schemes for high intensity field problems is still an area to be investigated.
In particular, with control problems in mind, see the recent works of
Raghunathan and Nest~\cite{Raghunathan2011,Raghunathan2012}.

In any case, once one has decided on a technique to describe the evolution of
molecules in the presence of intense laser pulses, it becomes possible to
attempt the inverse problem: given a desired behavior for the system, finding
what is the external perturbation that induces it. This type of problem is the
topic of ``control'' theories. Quantum optimal control theory
(QOCT)~\cite{Brif2010,Werschnik2007}, in particular, has been developed over
the last decades to answer the question of what are the best external handles
to manipulate a quantum system in order to achieve a predefined target.

QOCT has been applied to many systems for various purposes; in the molecular
physics field, most of the previous QOCT works have addressed the motion of
nuclear wave packets, on one or a few potential energy surfaces, in the
presence of femtosecond laser pulses. If a sufficient number of surfaces is
included in the model, and their non-adiabatic couplings properly treated,
this procedure is very precise. However, the computational cost of
pre-computing the surfaces with a good theory level, in addition the cost of
the propagation of the nuclear wave packets, makes it rather hard to apply.
If the classical approximation for the nuclei is good enough, a
mixed quantum-classical treatment is appropriate. A QOCT for mixed systems
would be necessary for these cases.

In a previous work~\cite{Krieger2011}, we have already studied the selective
bond breaking of molecules by making use of the Ehrenfest model. The method of
optimization, however, consisted of a simple gradient-free algorithm that does
not employ one of the essential results of QOCT: the computation of the
gradient of the target functional with respect to the field control
parameters. Recently, we have also presented the combination of QOCT with
TDDFT~\cite{Castro2012a}, which permits to directly control the electronic
motion, which occurs in the attosecond time scale, by employing TDDFT to
reduce the computational complexity of a full quantum electron dynamics. This
combination of QOCT with TDDFT did not consider the nuclear movement, and as
demonstrated in Ref.~\cite{Krieger2011} and more recently in
Ref.~\cite{Castro14}, this can only be valid if very short laser pulses are
considered.

In this work, we establish an optimal control framework for mixed systems
composed of quantum electrons and classical ions, modeled with the Ehrenfest
dynamics. The general framework is presented in
Section~\ref{section:oct-ehrenfest}, and it employs the fully correlated
many-electron wave functions. In order to establish a more manageable
practical formalism, we replace in Section~\ref{section:oct-ehrenfest-tddft}
that many-electron wave function by the single-particle orbitals that emerge
of TDDFT, combining the formalism presented in Ref.~\cite{Castro14} with that
of Section~\ref{section:oct-ehrenfest}. Finally, in Section~\ref{section:h2+}
the abstract general formulation is complemented with the simple example of
the H$_2^+$ molecule in the presence of a laser field.

\section{OCT for a general Ehrenfest system}
\label{section:oct-ehrenfest}

The state of a quantum-classical Ehrenfest system is specified by a set of
classical conjugated position and momenta variables $\lbrace
q_a,p_a\rbrace_{a}$, and a wave function $\Psi$. The dynamics of this system
is determined by a Hamiltonian $\hat{H}[q,p,u,t]$, which is a linear Hermitian
operator in the Hilbert space of wave functions, and simultaneosly depend on
the set of classical variables (we will denote $q$ to the full set of $q_a$
variables, and likewise $p$ to the set of $p_a$ variables). In general, the
Hamiltonian may also be time-dependent, and we consider its precise form to be
determined by a set of $M$ real parameters $u_1,\dots,u_M \equiv u$, which
constitute the ``control'' parameters. Usually, one separates out a
classical-only part, i.e. a part of the Hamiltonian that is a function of only
the classical variables:
\begin{equation}
\label{eq:qchamiltonian}
\hat{H}[q,p,u,t] = H_{\rm clas}[q,p,u,t]\hat{I} + \hat{H}_{\rm quantum}[q,p,u,t]\,.
\end{equation}
For example, one may include in $H_{\rm clas}$ the classical kinetic energy,
the interaction among the classical particles, or the action of external fields
on those classical particles. This choice, however, is somehow arbitrary, and the
inclusion or not of any of these purely classical terms into the quantum part merely
leads to different but physically irrelevant global time-dependent phase factors
in the wave function.

The dynamics is determined by this Hamiltonian through the following
set of ``Ehrenfest'' equations:
\begin{eqnarray}
\label{eq:ehr1}
\dot{q}_a(t) & = & \langle\Psi(t)\vert\frac{\partial \hat{H}}{\partial p_a}[q(t),p(t),u,t]\vert\Psi(t)\rangle
\\
\label{eq:ehr2}
\dot{p}_a(t) & = & -\langle\Psi(t)\vert\frac{\partial \hat{H}}{\partial q_a}[q(t),p(t),u,t]\vert\Psi(t)\rangle
\\
\label{eq:ehr3}
\dot{\Psi}(x,t) & = & -{\rm i} \hat{H}_{\rm quantum}[q(t),p(t),u,t]\Psi(x,t)\,,
\end{eqnarray}
complemented by a suitable set of initial conditions, $q_a(0) = q_{a0}\,,
p_a(0) = p_{a0}\,,\Psi(0) = \Psi_0$. The full set of quantum variables is
denoted collectively by $x$. Note that we assume that the Hamiltonian is
Hermitian, and therefore Eqs.~(\ref{eq:ehr1}) and (\ref{eq:ehr2}) can be
rewritten as:
\begin{eqnarray}
\dot{q}_a(t) & = & 
\frac{\partial \hat{H}_{\rm clas}}{\partial p_a}[q(t),p(t),u,t]+
\langle\Psi(t)\vert\frac{\partial \hat{H}_{\rm quantum}}{\partial p_a}[q(t),p(t),u,t]\vert\Psi(t)\rangle
\\
\dot{p}_a(t) & = & 
-\frac{\partial \hat{H}_{\rm clas}}{\partial q_a}[q(t),p(t),u,t]
-\langle\Psi(t)\vert\frac{\partial \hat{H}_{\rm quantum}}{\partial q_a}[q(t),p(t),u,t]\vert\Psi(t)\rangle
\end{eqnarray}

The purpose is to find a control $u$ that maximises some
objective, which could in principle be a functional of $p$, $q$, and
$\Psi$. This functional may have a ``terminal'' part (i.e. a
functional that depends on the state of the system at the final time
of the propagation, $T$), and a ``time-dependent part'' (i.e. a
functional of the full evolution of the system):
\begin{equation}
F[q,p,\Psi,u] = F^{\rm td}[q,p,\Psi,u] + F^{\rm term}[q(T),p(T),\Psi(T),u]\,.
\end{equation}
In order to simplify the following derivations, we will assume that
$F^{\rm td}$ is null, and only work with terminal targets.

The goal is to maximize the function:
\begin{equation}
G[u] = F[q[u], p[u], \Psi[u], u]\,.
\end{equation}
For this purpose, we must use Lagrange multipliers for each of the
dynamical variables: $\tilde{q}$, $\tilde{p}$ and $\chi$, and define
a function
\begin{equation}
  J[q,p,\tilde{q},\tilde{p},\Psi,\chi,u] = F[q,p,\Psi,u] + 
  L[q,p,\tilde{q},\tilde{p},\Psi,\chi,u]\,,
\end{equation}
where the Lagrangian functional $L$ is defined as:
\begin{eqnarray}
\nonumber
L[q,p,\tilde{q},\tilde{p},\Psi,\chi,u] = 
\\\nonumber
  - \sum_{a} \int_0^T \!\!\!\!\!\! {\rm d}t\; \tilde{p}_a(t) \left( \dot{q}_a(t) 
  - \langle \Psi(t) \vert \frac{\partial \hat{H}}{\partial p_a}[q(t),p(t),u,t]\vert\Psi(t)\rangle \right)
  \\\nonumber
  + \sum_{a} \int_0^T \!\!\!\!\!\! {\rm d}t\; \tilde{q}_a(t) 
      \left( \dot{p}_a(t) + \langle \Psi(t) \vert \frac{\partial \hat{H}}{\partial q_a}[q(t),p(t),u,t]\vert\Psi(t)\rangle \right)
  \\
  - 2{\rm Re} \int_0^T \!\!\!\!\!\! {\rm d}t\; 
  \langle \chi(t)\vert \frac{\rm d}{{\rm d}t} + {\rm i} \hat{H}_{\rm quantum}[q(t),p(t),u,t]\vert\Psi(t)\rangle\,.
\end{eqnarray}
This definition is designed to fulfill the following property:
The equations of motion (\ref{eq:ehr1}),(\ref{eq:ehr2}) and (\ref{eq:ehr3}) are retrieved by
taking functional derivatives with respect to the new variables and equating them to zero:
\begin{eqnarray}
\frac{\delta J}{\delta \tilde{q}_a(t)} =  
\frac{\delta L}{\delta \tilde{q}_a(t)} & = & 0\,,
\\
\frac{\delta J}{\delta \tilde{p}_a(t)} =  
\frac{\delta L}{\delta \tilde{p}_a(t)} & = & 0\,,
\\
\frac{\delta J}{\delta \chi^*(x,t)} = \,.
\frac{\delta L}{\delta \chi^*(x,t)} & = & 0\,.
\end{eqnarray}
These equations determine a map $u \rightarrow \lbrace q[u],p[u],\Psi[u]\rbrace$: the choice of 
a given control determines, through the equations of motion, the evolution of the system. In analogy, we may
obtain a set of equations of motion for the Lagrange multipliers:
we define them to be the result of setting the functional derivatives of $J$ with respect
to $q$, $p$ and $\Psi$ to zero. In order to compute these functional derivatives, it is better to rewrite
the Lagrangian function as:
\begin{eqnarray}
\nonumber
L[q,p,\tilde{q},\tilde{p},\Psi,\chi,u] =
   \sum_{a} \int_0^T \!\!\!\!\!\! {\rm d}t\; \left( \tilde{q}_a(t) \dot{p}_a(t) - \tilde{p}_a(t)\dot{q}_a(t) \right)
  \\\nonumber
   + \int_0^T \!\!\!\!\!\! {\rm d}t\; 
      \langle \Psi(t) \vert D_{\tilde{q}(t),\tilde{p}(t)} \hat{H}[q(t),p(t),u,t]\vert\Psi(t)\rangle
  \\
  - 2{\rm Re} \int_0^T \!\!\!\!\!\! {\rm d}t\; 
  \langle \chi(t)\vert \frac{\rm d}{{\rm d}t} + {\rm i} \hat{H}_{\rm quantum}[q(t),p(t),u,t]\vert\Psi(t)\rangle\,.
\end{eqnarray}
where the differential operator $D_{\tilde{q}(t),\tilde{p}(t)}$ is defined as:
\begin{equation}
D_{\tilde{q}(t),\tilde{p}(t)} = \sum_a \left( \tilde{q}_a(t)\frac{\partial}{\partial q_a} + 
                                         \tilde{p}_a(t)\frac{\partial}{\partial p_a} \right)
\end{equation}

The resulting equations of motion are:
\begin{eqnarray}
\nonumber
\dot{\tilde{q}}_a(t) & = &  \langle\Psi(t)\vert\;  
   D_{\tilde{q}(t),\tilde{p}(t)} \frac{\partial \hat{H}}{\partial p_a}[q(t),p(t),u,t] \;\vert\Psi(t)\rangle
\\\label{eq:ehr1-bis}
& & - 2 {\rm Re}\; {\rm i} \langle \chi(t) \vert 
     \frac{\partial \hat{H}_{\rm quantum}}{\partial p_a}[q(t),p(t),u,t]\vert\Psi(t)\rangle\,,
\\\nonumber
\dot{\tilde{p}}_a(t) & = & - \langle\Psi(t)\vert\; D_{\tilde{q}(t),\tilde{p}(t)} 
                            \frac{\partial \hat{H}}{\partial q_a}[q(t),p(t),u,t] \;\vert\Psi(t)\rangle
\\\label{eq:ehr3-bis}
& & + 2 {\rm Re}\; {\rm i} \langle \chi(t) \vert 
    \frac{\partial \hat{H}_{\rm quantum}}{\partial q_a}[q(t),p(t),u,t]\vert\Psi(t)\rangle\,,
\\\nonumber
\dot{\chi}(x,t) & = & -{\rm i} \hat{H}^\dagger_{\rm quantum}[q(t),p(t),u,t]\chi(x,t)
\\\label{eq:ehr5-bis}
& & + D_{\tilde{q}(t),\tilde{p}(t)} \hat{H}_{\rm quantum}[q(t),p(t),u,t]\Psi(x,t)\,,
\\
\label{eq:ehr2-bis}
\tilde{q}_a(T) & = & -\frac{\partial F^{\rm term}}{\partial p_a}[q(T),p(T),\Psi(T)]\,,
\\
\label{eq:ehr4-bis}
\tilde{p}_a(T) & = & \frac{\partial F^{\rm term}}{\partial q_a}[q(T),p(T),\Psi(T)]\,,
\\
\label{eq:ehr6-bis}
\chi(x, T) & = & \frac{\delta F^{\rm term}[q(T),p(T),\Psi(T)]}{\delta \Psi^*(x,T)}\,,
\end{eqnarray}
These equations establish the map $u \rightarrow \lbrace \tilde{q}[u],\tilde{p}[u],\chi[u]\rbrace$.
We may now proceed to compute the gradient of $G$. First, note that, for any
value of $u$, the Lagrangian function vanishes when we use the solution the mapped
arguments, i.e.:
\begin{equation}
L[q[u],p[u],\tilde{q}[u],\tilde{p}[u],\Psi[u],\chi[u],u] = 0\,,
\end{equation}
and therefore:
\begin{equation}
G[u] = J[q[u],p[u],\Psi[u],\tilde{q}[u],\tilde{p}[u],\chi[u],u]\,.
\end{equation}
The derivatives of $J$ with respect to any of its arguments (except
the explicit dependence on $u$) is zero, due to the manner in which we
have defined the maps $u \rightarrow q[u],\tilde{q}[u],\tilde{p}[u]$
and $u \rightarrow \Psi[u],\chi[u]$.  In consequence, the derivative
with respect to any of the parameters $u_m$ reduces on the right hand
side to only the explicit partial derivative, i.e.:
\begin{equation}
\label{eq:partialgpartialu}
\frac{\partial G}{\partial u_m}[u] = \left. 
\frac{\partial J[q,p,\Psi,\tilde{q},\tilde{p},\chi,u]}{\partial u_m}\right|_{
q=q[u],p=p[u],\Psi=\Psi[u],\tilde{q}=\tilde{q}[u],\tilde{p}=\tilde{p}[u],\chi=\chi[u]
}\,,
\end{equation}
which may be expanded to:
\begin{eqnarray}
\nonumber
  \frac{\partial G}{\partial u_m}[u] & = & \left.\frac{\partial F}{\partial u_m}[q,p,\Psi,u]
  \right|_{q=q[u](t),p=p[u](t),\Psi=\Psi[u](t)}
\\\nonumber
  & & 
   + \int_0^T\!\!\!\!\!\!{\rm d}t\; \langle \Psi[u](t)\vert D_{\tilde{q}[u](t),\tilde{p}[u](t)} 
   \frac{\partial \hat{H}}{\partial u_m}[q[u](t),p[u](t),u,t]
  \vert\Psi[u](t)\rangle
\\
  & & + 2 {\rm Im} \int_0^T\!\!\!\!\!\!{\rm d}t\; \langle 
     \chi[u](t) \vert \frac{\partial \hat{H}_{\rm quantum}}{\partial u_m}[q[u](t),p[u](t),u,t] \vert\Psi[u](t)\rangle\,.
\end{eqnarray}

\section{OCT for an Ehrenfest-TDDFT system}
\label{section:oct-ehrenfest-tddft}

In order to obtain the control equations for the case of Ehrenfest
dynamics in combination with TDDFT, some modification need to be done
to the previous scheme. In this section we derive the necessary
equations, which essentially consist of combining the formalism
developed in Ref.~\cite{Castro14}, with the one of previous section.

In TDDFT, the real interacting system of electrons is substituted by a
fictitious system of non-interacting electrons whose density is, by
definition, equal to the real one. Therefore, instead of one correlated wave
function we now have a Slater determinant. In order to simplify the formalism,
we will consider a spin-compensated system with an even number $N$ of
electrons doubly occupying $N/2$ orbitals $\varphi_i$.  The one-body density
of this Slater determinant is given by;
\begin{equation}
n(\vec{r},t) \equiv n_t(\vec{r}) = \sum_{j=1}^{N/2} 2\vert \varphi_j(\vec{r},t)\vert^2\,.
\end{equation}
The one-particle Hamiltonian that governs the motion of the non-interacting
electrons is a functional of this density, and is given the name of
``Kohn-Sham Hamiltonian''. In this context, it also depends on the classical
variables $(q,p)$, and on the control parameters $u$. The full Hamiltonian
that takes the place of the one in Eq.~(\ref{eq:qchamiltonian}) may in this case be
written as:
\begin{equation}
\hat{H}[q,p,u,t] = H_{\rm clas}[q,p,u,t]\hat{I} + \sum_{i=1}^N\hat{H}_{\rm KS}^{(i)}[q,p,n_t,u,t]\,,
\end{equation}
where $\hat{H}_{\rm KS}^{(j)}[q,p,n_t,u,t]$ is the one-particle Kohn-Sham Hamiltonian, acting on particle $i$.
Note that this Hamiltonian depends on the electronic density at time $t$, $n_t$. This is
in fact an approximation - the \emph{adiabatic} approximation - which we take here
because it simplifies the notation of the results given below, and because the vast majority
applications of TDDFT up to now use it.

The corresponding equations of motion are:
\begin{eqnarray}
\nonumber
\dot{q}_a(t) & = & \frac{\partial H_{\rm clas}}{\partial p_a}[q(t),p(t),u,t] + 
\\
& & 
  \sum_{j=1}^{N/2} 2\langle\varphi_j(t)\vert\frac{\partial 
  \hat{H}_{\rm KS}}{\partial p_a}[q(t),p(t),u,t]\vert\varphi_j(t)\rangle
\label{eq:ieehr1}
\\\nonumber
\dot{p}_a(t) & = & 
-\frac{\partial H_{\rm clas}}{\partial q_a}[q(t),p(t),u,t]
\\
& & 
-\sum_{j=1}^{N/2} 2\langle\varphi_j(t)\vert\frac{\partial 
  \hat{H}_{\rm KS}}{\partial q_a}[q(t),p(t),u,t]\vert\varphi_j(t)\rangle
\label{eq:ieehr2}
\\
\label{eq:ieehr3}
\dot{\varphi}_j(\vec{r},t) & = & -{\rm i} \hat{H}_{\rm KS}[q(t),p(t),u,t]\varphi_j(\vec{r},t)\,,
\end{eqnarray}
Here we have assumed the following: the derivatives $\displaystyle \frac{\partial 
  \hat{H}_{\rm KS}}{\partial q_a}$ and $\displaystyle \frac{\partial 
  \hat{H}_{\rm KS}}{\partial p_a}$ do not depend on the electronic density. The reason is that 
the density is included
in the Kohn-Sham Hamiltonian through the Hartree and exchange-correlation potentials, which
do not depend (explicitly) on the classical variables. In fact, it will later
be useful to split the Kohn-Sham Hamiltonian in the following manner:
\begin{equation}
\hat{H}_{\rm KS}[q(t),p(t),n_t,u,t] = \hat{H}^0_{\rm KS}[q(t),p(t),u,t] + \hat{V}_{\rm Hxc}[n_t]
\end{equation}

The previous Eqs.~(\ref{eq:ieehr1}), (\ref{eq:ieehr2}) and (\ref{eq:ieehr3}) determine the
evolution of the system, given a choice for the control parameters: $u
\rightarrow \lbrace q[u],p[u],\varphi[u]\rbrace$. The goal, as in
previous section, is to maximize a function $G$ defined in terms of a
functional of the system behaviour:
\begin{equation}
G[u] = F^{\rm term}[q[u](T),p[u](T),\varphi[u](T), u]\,.
\end{equation}
Once again, we have assumed that this target depeds only on the final state of the system. The
computation of the gradient of this function proceeds as in previous section, by defining
a suitable extended functional, depending on a set of Lagrange multipliers $\tilde{q},\tilde{p},\chi$:
\begin{equation}
J[q,p,\varphi,\tilde{p},\tilde{p},\chi,u] = F^{\rm term}[q,p,\varphi, u] +
L[q,p,\varphi,\tilde{q},\tilde{p},\chi,u]
\end{equation}
with the help of the following Lagragian:
\begin{eqnarray}
\nonumber
L[q,p,\varphi,\tilde{q},\tilde{p},\chi,u] = 
\\\nonumber
  - \sum_{a} \int_0^T \!\!\!\!\!\! {\rm d}t\; \tilde{p}_a(t) \left( \dot{q}_a(t) -
  \frac{\partial H_{\rm clas}}{\partial p_a}[q(t),p(t),u,t] 
  \phantom{\frac{\partial \hat{H}_{\rm KS}}{\partial p_a}} 
  \right.
\\\nonumber
  \left.
  \phantom{- \sum_{a} \int_0^T \!\!\!\!\!\! {\rm d}t\; \tilde{p}_a(t)}
  - \sum_{j=1}^{N/2} 2\langle\varphi_j(t)\vert\frac{\partial 
   \hat{H}_{\rm KS}}{\partial p_a}[q(t),p(t),u,t]\vert\varphi_j(t)\rangle
  \right)
\\\nonumber
  + \sum_{a} \int_0^T \!\!\!\!\!\! {\rm d}t\; \tilde{q}_a(t) 
      \left( \dot{p}_a(t) + 
\frac{\partial H_{\rm clas}}{\partial q_a}[q(t),p(t),u,t] 
\phantom{\frac{\partial \hat{H}_{\rm KS}}{\partial q_a}}
\right.
\\\nonumber
\left.
  \phantom{+ \sum_{a} \int_0^T \!\!\!\!\!\! {\rm d}t\; \tilde{q}_a(t) }
+\sum_{j=1}^{N/2} 2\langle\varphi_j(t)\vert\frac{\partial 
  \hat{H}_{\rm KS}}{\partial q_a}[q(t),p(t),u,t]\vert\varphi_j(t)\rangle \right) \,.
  \\
  - 2{\rm Re} \sum_{j=1}^{N/2}\int_0^T \!\!\!\!\!\! {\rm d}t\; 
  \langle \chi_j(t)\vert \frac{\rm d}{{\rm d}t} + {\rm i} \hat{H}_{\rm KS}[q(t),p(t),n_t,u,t]\vert\varphi_j(t)\rangle\,.
\end{eqnarray}
The functional derivatives of $J$ with respect to the new variables $\tilde{q},\tilde{p}$ and $\chi$, set to zero, lead to the equations of motion of the system. In order to get the equations of motion for the new
variables, we must compute and set to zero the functional derivatives of $J$ with respect to the orginal
system variables. In order to do this, it is helpful to rewrite the Lagrangian as:
\begin{eqnarray}
\nonumber
L[q,p,\tilde{q},\tilde{p},\varphi,\chi,u] = 
\\\nonumber
   \sum_{a} \int_0^T \!\!\!\!\!\! {\rm d}t\; \left( \tilde{q}_a(t)\dot{p}_a(t) - \tilde{p}_a(t)  \dot{q}_a(t) \right)
\\\nonumber
  + \int_0^T \!\!\!\!\!\! {\rm d}t\; D_{q(t),p(t)}
  H_{\rm clas}[q(t),p(t),u,t] 
\\\nonumber
  + \sum_{j=1}^{N/2} \int_0^T \!\!\!\!\!\! {\rm d}t\; 2\langle\varphi_j(t)\vert 
   D_{q(t),p(t)}\hat{H}_{\rm KS}[q(t),p(t),u,t]\vert\varphi_j(t)\rangle
  \\\nonumber
  - 2{\rm Re} \sum_{j=1}^{N/2}\int_0^T \!\!\!\!\!\! {\rm d}t\; 
  \langle \chi_j(t)\vert \frac{\rm d}{{\rm d}t} + {\rm i} \hat{H}^0_{\rm KS}[q(t),p(t),n_t,u,t]\vert\varphi_j(t)\rangle\,.
\\
 + L^{\rm Hxc}[\varphi,\chi]\,.
\end{eqnarray}
In this expression, we have separated out the part that contains the non-linear Hartree, exchange
and correlation terms:
\begin{equation}
L^{\rm Hxc}[\varphi,\chi] = 
  - 2{\rm Re}\; {\rm i}\sum_{j=1}^{N/2}\int_0^T \!\!\!\!\!\! {\rm d}t\; 
  \langle \chi_j(t)\vert  \hat{V}_{\rm Hxc}[n_t]\vert\varphi_j(t)\rangle\,.
\end{equation}
The functional derivatives of this term with respect to the Kohn-Sham orbitals are:
\begin{eqnarray}
\nonumber
\frac{\delta L^{\rm Hxc}}{\delta \varphi^*_l(\vec{r},t)} & = &
 \varphi_l(\vec{r},t) 4 {\rm Im}\; \sum_{j=1}^{N/2}\int\!{\rm d}^3r' 
\chi^*_j(\vec{r},t)
f_{\rm Hxc}[n_t](\vec{r},\vec{r}')
\varphi_j(\vec{r}',t) 
\\
& & 
+ {\rm i} v_{\rm Hxc}[n_t](\vec{r})\chi_l(\vec{r},t)\,.
\end{eqnarray}
If we now define the following set of operators:
\begin{equation}
\hat{K}_{lj}[\varphi(t)]\psi (\vec{r}) = -4{\rm i} \varphi_l(\vec{r},t) {\rm Im}\;
\int\!{\rm d}^3r' \psi(\vec{r}) f_{\rm Hxc}[n_t](\vec{r},\vec{r}') \varphi_j(\vec{r},t)\,,
\end{equation}
we may rewrite the previous functional derivative as:
\begin{equation}
\frac{\delta L^{\rm Hxc}}{\delta \varphi^*_l(\vec{r},t)}  = 
 {\rm i} \sum_{j=1}^{N/2}\hat{K}_{lj}[\varphi(t)]\chi_j(\vec{r},t)
+ {\rm i} v_{\rm Hxc}[n_t](\vec{r})\chi_l(\vec{r},t)\,.
\end{equation}

And the resulting equations of motion are:
\begin{eqnarray}
\nonumber
\dot{\tilde{q}}_a(t) & = & \frac{\partial}{\partial p_a} D_{q(t),p(t)}H_{\rm class}[q(t),p(t),u,t]
\\\nonumber
& &
+ \sum_{j=1}^{N/2}2\langle\varphi_j(t)\vert \frac{\partial}{\partial p_a} D_{q(t),p(t)} \hat{H}_{\rm KS}[q(t),p(t),u,t]\vert
\varphi_j(t)\rangle
\\
& &
- 2{\rm Re}\;{\rm i}\sum_{j=1}^{N/2} \langle \chi_j(t)\vert \frac{\partial}{\partial p_a} 
  \hat{H}_{\rm KS}[q(t),p(t),u,t]\vert
  \varphi_j(t)\rangle\,, 
\\\nonumber
\dot{\tilde{p}}_a(t) & = & - \frac{\partial}{\partial q_a} D_{q(t),p(t)}H_{\rm class}[q(t),p(t),u,t]
\\\nonumber
& &
- \sum_{j=1}^{N/2}2\langle\varphi_j(t)\vert \frac{\partial}{\partial p_a} D_{q(t),p(t)} \hat{H}_{\rm KS}[q(t),p(t),u,t]\vert
\varphi_j(t)\rangle
\\
& &
+ 2{\rm Re}\;{\rm i}\sum_{j=1}^{N/2} \langle \chi_j(t)\vert \frac{\partial}{\partial p_a} 
  \hat{H}_{\rm KS}[q(t),p(t),u,t]\vert
  \varphi_j(t)\rangle\,, 
\\\nonumber
\dot{\chi}_j(\vec{r},t) & = & -{\rm i}\hat{H}_{\rm KS}[q(t),p(t),n_t,u,t]\chi_j(\vec{r},t)
\\\nonumber
& & 
- {\rm i}\sum_{k=1}^{N/2} K_{jk}[\varphi(t)]\chi_k(\vec{r},t)\,,
\\
& & 
+ 2 \sum_{j=1}^{N/2} D_{q(t),p(t)} H_{\rm KS}[q(t),p(t),u,t] \varphi_j(\vec{r},t)
\\
\tilde{q}_a(T) & = & - \frac{\partial F^{\rm term}}{\partial p_a}[q(T),p(T),\varphi(T)]
\\
\tilde{p}_a(T) & = & \frac{\partial F^{\rm term}}{\partial q_a}[q(T),p(T),\varphi(T)]
\\
\chi_j(\vec{r},T) & = & \frac{\delta F^{\rm term}}{\delta \varphi^*(\vec{r},T)}
\end{eqnarray}
These equations establish the map $u\rightarrow \chi[u]$. This is the
ingredient needed to compute the gradient of $G$, which, in analogy to the
Eq.~(\ref{eq:partialgpartialu}) obtained in the previous section, is given by:
\begin{equation}
\frac{\partial G}{\partial u_m}[u] = \left. 
\frac{\partial J[q,p,\varphi,\tilde{q},\tilde{p},\chi,u]}{\partial u_m}\right|_{
q=q[u],p=p[u],\varphi=\varphi[u],\tilde{q}=\tilde{q}[u],\tilde{p}=\tilde{p}[u],\chi=\chi[u]
}\,,
\end{equation}


\section{The H$_2^+$ molecule}
\label{section:h2+}

We finish by particularizing the previous rather abstract formalism to
the case of the simplest of molecules, H$_2^+$, composed of two
protons and one electron, in the presence of an electric field. To
simplify even further, so that the resulting equations are as clear as
possible, we will reduce the number of classical degrees of freedom to
only one (the internuclear distance). To achieve this, we will work in
the reference frame of the nuclear center of mass, neglect the
inertial force due to its acceleration, and we will assume cylindrical
symmetry along the molecular axis.

The quantum-classical
Hamiltonian is given by:
\begin{eqnarray}
\nonumber
\hat{H} & = &  
\frac{1}{2M}\vec{P}_1^2 + 
\frac{1}{2M}\vec{P}_1^2 + 
\frac{1}{2}\hat{\vec{p}}^2 
\\\nonumber
&  & 
+ w(\vec{R}_1-\vec{R}_2)
- w(\hat{\vec{r}}-\vec{R}_1)
- w(\hat{\vec{r}}-\vec{R}_2)
\\
& & 
- \varepsilon(u,t)\vec{\pi}\cdot\vec{R}_1
- \varepsilon(u,t)\vec{\pi}\cdot\vec{R}_2
+ \varepsilon(u,t)\vec{\pi}\cdot\hat{\vec{r}}\,.
\end{eqnarray}
In this equation, $(\vec{R}_1,\vec{P}_1)$ and $(\vec{R}_2,\vec{P}_2)$ are the
position and momentum pairs of the two (classical) protons, and
$(\hat{\vec{r}},\hat{\vec{p}})$ is the position and momentum operator pair of
the electron. $M$ is the proton mass in atomic units, and $w$ is the
particle-particle interaction function (the proton-proton and electron-proton
interactions are identical, except for the opposite sign). The last terms
are the interaction of the particles with an electric field
$\varepsilon(u,t)\vec{\pi}$ in the dipole approximation.

It is convenient to transform the classical variables into the nuclear
center-of-mass and relative particle coordinates:
\begin{eqnarray}
\vec{R} & = & \vec{R}_1 - \vec{R}_2\,,
\\
\vec{R}_{\rm CM} & = & \frac{1}{2}(\vec{R}_1 + \vec{R}_2)\,.
\end{eqnarray}
The Hamiltonian changes into:
\begin{eqnarray}
\nonumber
\hat{H} & = & 
\frac{1}{2M_{\rm CM}}\vec{P}_{\rm CM}^2 +
\frac{1}{2\mu}\vec{P}^2 
+ w(\vec{R})
- \varepsilon(u,t)\vec{\pi}\cdot\vec{R}_{\rm CM}
\\
& & 
+ \frac{1}{2}\hat{\vec{p}}^2 
+ \hat{V}(\hat{\vec{r}};\vec{R}_{\rm CM},\vec{R})\,,
\end{eqnarray}
where:
\begin{eqnarray}
\nonumber
\hat{V}(\hat{\vec{r}};\vec{R}_{\rm CM},\vec{R}) & = & - w(\hat{\vec{r}}-\vec{R}_{\rm CM} - \frac{1}{2}\vec{R})
- w(\hat{\vec{r}}-\vec{R}_{\rm CM} + \frac{1}{2}\vec{R})
\\
& & + \varepsilon(u,t)\vec{\pi}\cdot (\hat{\vec{r}} - \vec{R}_{\rm CM})\,.
\end{eqnarray}
The mass of the nuclear center of mass $M_{\rm CM}$ is $2M$, whereas the
reduced mass $\mu$ is $M/2$. This full Hamiltonian can be conveniently split
into a classical and a quantum part as:
\begin{equation}
H_{\rm clas} = \frac{1}{2M_{\rm CM}}\vec{P}_{\rm CM}^2 + 
\frac{1}{2\mu}\vec{P}^2 + w(\vec{R}) - \varepsilon(u,t)\vec{\pi}\cdot
\vec{R}_{\rm CM}\,.
\end{equation}
\begin{eqnarray}
\nonumber
\hat{H}_{\rm quantum} & = &  
\frac{1}{2}\hat{\vec{p}}^2 + \hat{V}(\hat{\vec{r}};\vec{R}_{\rm CM},\vec{R})
\end{eqnarray}
By noticing that
\begin{equation}
\hat{V}(\hat{\vec{r}}+\vec{R}_{\rm CM};\vec{R}_{\rm CM},\vec{R}) = 
\hat{V}(\hat{\vec{r}};\vec{0},\vec{R})\,,
\end{equation}
it becomes clear that some simplification is to be expected if we move to the
reference system of the center of mass. Schr{\"{o}}dinger's equation for the
electron is:
\begin{equation}
i\frac{\rm d}{{\rm d}t} \vert\psi(t)\rangle = 
\left[
\frac{1}{2}\hat{\vec{p}}^2 + \hat{V}(\hat{\vec{r}};\vec{R}_{\rm CM},\vec{R})
\right] \vert\psi(t)\rangle\,,
\end{equation}
but we may instead perform a unitary transformation in the form:
\begin{equation}
\vert\Psi(t)\rangle = \hat{U}[\vec{R}_{\rm CM}(t)]\vert\psi(t)\rangle\,,
\end{equation}
where
\begin{equation}
\hat{U}[\vec{R}_{\rm CM}(t)] =
\exp(i\vec{R}_{\rm CM}(t)\cdot\hat{\vec{p}}-i\dot{\vec{R}}_{\rm CM}(t)\cdot\hat{\vec{r}}
+\frac{i}{2}\int_0^t\!{\rm d}\tau \dot{\vec{R}}_{\rm CM}^2(\tau))\,.
\end{equation}
The corresponding Schr{\"{o}}dinger's equation for this transformed state is~\cite{Rosen1972,Takagi1991}:
\begin{equation}
i\frac{\rm d}{{\rm d}t} \vert\Psi(t)\rangle = 
\left[
\frac{1}{2}\hat{\vec{p}}^2 + \hat{V}(\hat{\vec{r}};\vec{0},\vec{R})
\right] \vert\Psi(t)\rangle + \ddot{\vec{R}}_{\rm CM}(t)\cdot\hat{\vec{r}}\vert\Psi(t)\rangle\,.
\end{equation}
Note the presence of an \emph{inertial} term, due to the acceleration of the
nuclear center of mass. If we assume this term to be small (an assumption
which is based on the heavy weight of the nuclei), the previous equation is
completely decoupled from the center of mass variable, and we may write:
\begin{equation}
i\frac{\rm d}{{\rm d}t} \vert\Psi(t)\rangle = 
\left[
\frac{1}{2}\hat{\vec{p}}^2 
- w(\hat{\vec{r}} - \frac{1}{2}\vec{R})
- w(\hat{\vec{r}} + \frac{1}{2}\vec{R}) 
+ \varepsilon(u,t)\vec{\pi}\cdot\hat{\vec{r}}
\right] \vert\Psi(t)\rangle\rangle\,.
\end{equation}

The equation of motion for the relative particle can then also be exactly written
without the presence of the center of mass variables:
\begin{eqnarray}
\nonumber
\frac{\rm d}{{\rm d}t}\vec{P}(t) & = & -\nabla w(\vec{R})-\langle \psi(t)\vert
\nabla_{\vec{R}} V(\hat{\vec{r}}(t),\vec{R}_{\rm CM}(t),\vec{R}(t))\vert\psi(t)\rangle
\\\nonumber
& = & 
-\nabla w(\vec{R})-\langle \Psi(t)\vert
\nabla_{\vec{R}} V(\hat{\vec{r}}(t),\vec{0},\vec{R}(t))\vert\Psi(t)\rangle
\\
& = & 
-\nabla w(\vec{R})-\langle \Psi(t)\vert
\frac{1}{2}\nabla w(\hat{\vec{r}}-\frac{1}{2}\vec{R})-
\frac{1}{2}\nabla w(\hat{\vec{r}}-\frac{1}{2}\vec{R})
\vert\Psi(t)\rangle
\end{eqnarray}

The last two equations can be considered to be derived from the following
quantum and classical Hamiltonians, that consider the relative particle only:
\begin{equation}
H_{\rm clas} = \frac{1}{2\mu}\vec{P}^2 + w(\vec{R})\,,
\end{equation}
\begin{equation}
\hat{H}_{\rm quantum} =
\frac{1}{2}\hat{\vec{p}}^2 
- w(\hat{\vec{r}}- \frac{1}{2}\vec{R})
- w(\hat{\vec{r}}+ \frac{1}{2}\vec{R})
+ \varepsilon(u,t)\vec{\pi}\cdot\hat{\vec{r}}\,.
\end{equation}
We may simplify the problem further by considering the existence of
cylindrical symmetry around the molecular axis, which requires that the
electrical field is directed in that direction: $\vec{\pi}=\vec{z}$, assuming
that the molecular axis is the $z$-direction. If the initial momentum is zero
(or is also parallel to the $z$ axis), and the initial electronic wave function
is cylindrically symmetric, then this symmetry will be preserved and we need
only take care of the $z$ component $q=\vec{R}\cdot\vec{z}$, and its
corresponding momentum $p=\vec{P}\cdot\vec{z}$. Therefore, if we define:
\begin{equation}
v(q,\hat{\vec{r}}) = - w(\hat{\vec{r}}- \frac{1}{2}q\vec{z})
- w(\hat{\vec{r}}+ \frac{1}{2}q\vec{z})\,,
\end{equation}
we may finally describe the system with the following Hamiltonians:
\begin{equation}
H_{\rm clas} = \frac{1}{2\mu}p^2 + w(q)\,,
\end{equation}
\begin{equation}
\hat{H}_{\rm quantum} =
\frac{1}{2}\hat{\vec{p}}^2 
+ v(q,\hat{\vec{r}})
+ \varepsilon(u,t)\vec{\pi}\cdot\hat{\vec{r}}\,.
\end{equation}
The corresponding equations of motion are:
\begin{eqnarray}
\dot{p}(t)  & = & -w'(q(t))
- \langle\Psi(t)\vert
\frac{\partial v}{\partial q}(q(t),\hat{\vec{r}})
\vert\Psi(t)\rangle\,,
\\
\dot{q}(t) & = & \frac{1}{\mu}{p}(t)\,,
\\
i\frac{\rm d}{{\rm d}t}\vert\Psi(t)\rangle & = & 
\left[
\frac{1}{2}\hat{\vec{p}}^2 
+ v(q(t),\hat{\vec{r}})
+ \varepsilon(u,t)\vec{\pi}\cdot\hat{\vec{r}}
\right]
\vert\Psi(t)\rangle\,.
\end{eqnarray}
Once we have a dynamical system clearly defined, we can proceed to 
pose and solve optimization problems.
For example, one may wish to find a laser
pulse that dissociates the molecule. This can be formulated by requiring the
maximization of the relative coordinate $q$ at the final time of the
propagation. One may therefore define, for example:
\begin{equation}
F^{\rm term}[q(T),p(T),u] = q^2(T)\,,
\end{equation}
so that:
\begin{equation}
G[u] = q^2[u](T)\,.
\end{equation} 
We may now directly apply the expressions obtained in
Section~\ref{section:oct-ehrenfest}. The result is the following: The equation
for the gradient of function $G$ is:
\begin{equation}
  \frac{\partial G}{\partial u_m}[u] =
   2 {\rm Im} \int_0^T\!\!\!\!\!\!{\rm d}t\; \langle 
     \chi[u](t) \vert\;
\frac{\partial \varepsilon}{\partial u_m}(u,t)\vec{\pi}\cdot\hat{\vec{r}}
\;\vert\Psi[u](t)\rangle\,.
\end{equation}
In order to compute this expression one needs the ``Lagrange multiplier'' wave function
$\chi$, which can be obtained by backwards propagation of its equation of
motion. This equation, along with the also necessary equations for the
other auxiliary Lagrange multiplier variables $\tilde{q},\tilde{p}$, is:
\begin{eqnarray}
\nonumber
\dot{\tilde{q}}(t) & = &  \frac{\tilde{p}(t)}{\mu}
\\\nonumber
\dot{\tilde{p}}(t) & = & - \langle\Psi(t)\vert\; 
\tilde{q}(t)\frac{\partial^2 v}{\partial q^2}(q(t),\hat{\vec{r}}) +
\tilde{q}(t)w''(q(t))
\;\vert\Psi(t)\rangle
\\
& & + 2 {\rm Re}\; {\rm i} \langle \chi(t) \vert 
\frac{\partial v}{\partial q}(q(t),\hat{\vec{r}})
\vert\Psi(t)\rangle\,,
\\\nonumber
\dot{\chi}(x,t) & = & -{\rm i} \hat{H}^\dagger_{\rm quantum}[q(t),p(t),u,t]\chi(x,t)
\\
& & 
\left[
\tilde{q}(t) \frac{\partial v}{\partial
  q}(q(t),\hat{\vec{r}}) + \tilde{q}(t) w'(q(t)) + \frac{1}{\mu}\tilde{p}(t)p(t)
\right]
\Psi(x,t)\,,
\end{eqnarray}
And, to conclude, the final-time conditions are:
\begin{eqnarray}
\tilde{q}(T) & = & 0\,,
\\
\tilde{p}(T) & = & 2q(T)\,,
\\
\chi(x, T) & = & 0\,.
\end{eqnarray}












\section{Conclusions}

Ehrenfest MD based on TDDFT is a computationally practical model, as
demonstrated in the past by numerous studies. The evolution of molecular
systems in the presence of laser fields can be simulated in reasonable times,
depending, of course, on the size of the molecular system and on the required
propagation time. One may then wonder whether it is also possible to perform
optimization calculations with this model: this means, in the context of
molecules irradiated with laser pulses, the calculation of those pulse shapes
that induce an optimal behavior of the system, as defined by a given target
functional.

While control theory in the context of engineering problems (obviously
addressing classical problems), and QOCT are already mature
disciplines, there has been no attempt to extend optimal control
theory to quantum-classical models. In this work, we have
presented the fundamental equations of an optimal control theory for
systems containing both quantum electrons and classical ions. In
particular, the model of choice has been Ehrenfest dynamics, a
non-adiabatic variant of molecular dynamics. The general formulation,
that needs the fully correlated many-electron wave function, can be
simplified by making use of TDDFT. In this case, the optimal control
equations require some modifications that we have also provided.

The key equations that we have derived are those that permit to compute the
gradient of the target function with respect to the optimizing
parameters. Armed with this gradient, one can use any of the various
non-linear optimization algorithms available. In essence, the required
computations amount to the forwards propagation of the system itself, along
with a backwards propagation of an auxiliary system. The computational
complexity of this backwards propagation is similar to the complexity of the
forwards propagation, and therefore one may conclude that the optimization is
feasible as long as the propagation of the initial model is also
feasible. Work towards the numerical implementation of these ideas is in
progress. For this purpose we will use the optimal control capabilities
already implemented in the octopus code~\cite{Marques2003,Castro2006}, which
has been used for electronic-only control problems in various previous works,
e.g. Refs~\cite{Rasanen2007,Kammerlander2011a,Krieger2011,Rasanen2012,Castro14,Blasi2013}.

\section*{References}

\bibliographystyle{unsrt}
\bibliography{/home/alberto/Documents/library.bib,paper.bib}

\end{document}